\documentclass[a4paper]{aipproc}

\layoutstyle{6x9}

\begin{document}

\title
      [Microscopic description of $\eta$-photoproduction
on light nuclei.]
      {Microscopic description of $\eta$-photoproduction \\
on light nuclei.}


\author{V. B. Belyaev}{
  address={Joint Institute  for Nuclear Research, Dubna,
  141980, Russia}
}

\iftrue
\author{N. V. Shevchenko}{
  address={Joint Institute  for Nuclear Research, Dubna,
  141980, Russia}
}

\author{S. A. Rakityansky}{
  address={Physics Dept., University of South Africa,
         P.O. Box 392, Pretoria 0003, South Africa}
}

\author{W.~Sandhas}{
  address={Physikalisches Institut, Universit\H{a}t Bonn,
  D-53115 Bonn, Germany}
}

\author{S. Sofianos}{
  address={Physics Dept., University of South Africa,
         P.O. Box 392, Pretoria 0003, South Africa}
}
\fi

\copyrightyear {2001}

\begin{abstract}
A microscopic four-body description of near-threshold coherent
photoproduction of the $\eta$ meson on the ($3N$)-nuclei is
given. The photoproduction cross-section is calculated using the
Finite Rank Approximation (FRA) of the nuclear Hamiltonian.  The
results indicate that the final state interaction of the $\eta$
meson with the residual nucleus plays an important role in the
photoproduction process. Sensitivity of the results to the
choice of the $\eta N$ $T$-matrix is investigated.  The
importance of obeying the condition of $\eta N$ unitarity is
demonstrated.
\end{abstract}

\date{\today}

\maketitle

The high level of reliability of the modern few-body theory
provides the means for making conclusions about underlying
two-body interactions from experimental data on relevant
few-body processes. The purpose of this report is to present the
results of few-body calculations concerning the coherent
$\eta$-photoproduction on the tritium and $^3$He targets,
\begin{equation}
\label{reactions}
   {}^3{\rm H}(\gamma, \eta){}^3{\rm H}\qquad
   \mbox{and}\qquad
   {}^3{\rm He}(\gamma, \eta){}^3{\rm He}\ ,
\end{equation}
from which some conclusions about the $T$-matrices describing
the elastic $\eta N$ scattering and the photoproduction process
$N(\gamma, \eta)N$ could be made.  To the best of our knowledge,
no experimental data on such coherent reactions has been
published yet.

To describe the few-body dynamics of this reaction, we employ
the method based on the Finite-Rank Approximation (FRA)
\cite{FRA} of the nuclear Hamiltonian $H_A$.  This approximation
consists in neglecting the continuous spectrum in the spectral
expansion
\begin{equation}
\label{expansion}
   H_A={\cal E}_0|\psi_0\rangle\langle\psi_0|+
       \mbox{continuum}
\end{equation}
of this Hamiltonian.  Physically, this means that we exclude the
processes of (virtual) excitations of the nucleus during its
interaction with the $\eta$ meson. It is clear that the stronger
the nucleus is bound, the smaller is the contribution from such
processes to the elastic $\eta A$ scattering.  By comparing with
the results of the exact Faddeev calculations, it was
shown\cite{AGSetad} that even for $\eta d$ scattering (with
weakest nuclear binding) the FRA method works reasonably well,
which implies that we can obtain sufficiently accurate results
applying this method to $\eta{}^3$H and $\eta{}^3$He scattering.

To include a photon into the FRA formalism, we follow the same
procedure as in Ref.\cite{Ours3} where the coherent
$\eta$-photoproduction on deuteron was treated in the framework
of the exact AGS equations and the photon was introduced by
considering the $\eta N$ and $\gamma N$ states as two different
channels of the same system. This implies that the $T$-operator
describing the $\eta N$ interaction, should be replaced by $2
\times 2$ matrix, viz.
\begin{equation}
\label{matrix} t_{\eta N} \to \left(
\begin{array}{cc}
 t^{\gamma \gamma} & t^{\gamma \eta} \\
 t^{\eta \gamma}   & t^{\eta \eta}
\end{array}
\right)\ ,
\end{equation}
where $t^{\gamma \gamma}$ describes the Compton scattering,
$t^{\eta \gamma}$ the photoproduction process, and $t^{\eta
\eta}$ the elastic $\eta N$ scattering. All calculations was
performed in the first order on electromagnetic interaction.

The problem of constructing $\eta N$ potential or directly the
corresponding $T$-matrix $t^{\eta\eta}$ has no unique solution
since the only experimental information we have consists of the
two complex numbers, namely, position of the $S_{11}$-resonance
pole $E_0-i\Gamma/2$ and the $\eta N$ scattering length $a_{\eta
N}$. In the present work, we use three different versions of
$t^{\eta\eta}$ all of which have the same separable form
\begin{equation}
    t^{\eta\eta}(k',k;z) = g(k') \, \tau(z) \, g(k)\ ,
\label{tetan}
\end{equation}
with the same formfactors $g(k)=(k^2+\alpha^2)^{-1}$ (where
$\alpha = 3.316$\,fm$^{-1}$, see Ref.~\cite{alpha}) but with
different versions of the propagator $\tau(z)$. The version I is
motivated by the dominance of the $S_{11}$ resonance at the
near-threshold energies and has simple Breit-Wigner form
\begin{equation}
    \tau(z) = \frac {\lambda}{z - E_0 + i\Gamma/2}\ ,
\label{tau1}
\end{equation}
which guaranties that the $t^{\eta\eta}$ has the resonance pole
at $z=E_0-i\Gamma/2$ (with $E_0=1535\,{\rm MeV}-(m_N+m_\eta)$
and $\Gamma=150\,{\rm MeV}$ see Ref. \cite{PDG}). In this
version of $\tau(z)$  the strength parameter $\lambda$ is chosen
to reproduce the $\eta N$ scattering length $a_{\eta N} = (0.55
+ i 0.30)\,{\rm fm}$.

An alternative way (version II) of constructing the two-body
$T$-matrix $t^{\eta\eta}$ is to solve the corresponding
Lippmann-Shwinger equation with an appropriate separable
potential having the same form-factors $g(k)$.  However, a
one-term separable $T$-matrix obtained in this way, does not
have a pole at $z=E_0-i\Gamma/2$.  To recover the resonance
behaviour in this case, we use the trick suggested in
Ref.~\cite{Deloff}, namely, we use a potential with an
energy-dependent strength, which resulted in
\begin{equation}
    \tau(z) = -\frac{4 \pi \alpha^3}{\mu_{\eta N}}\cdot
    \frac{\Lambda (\zeta-z) + C \zeta}
    {\zeta-z-\left[\Lambda (\zeta-z)
    + C \zeta\right] / (1 - i\sqrt{2 z \mu_{\eta N}} /\alpha)^2}\ ,
\label{tdeloff}
\end{equation}
where the constants $\Lambda$, $C$, and $\zeta$ are chosen in
such way that the corresponding scattering amplitude reproduces
the same (as for version I) scattering length $a_{\eta N}$ and
has a pole at $z=E_0-i\Gamma/2$.  Moreover, it is consistent
with the condition of the two-body unitarity because it obeys
the Lippmann-Schwinger equation.

The version III has the same functional form as version I, but
different value of $\lambda$ which is now fixed using the
condition of the $\eta N$ unitarity, namely, $(1-2\pi
it^{\eta\eta})(1-2\pi it^{\eta\eta})^+=1$.  The resulting
$t^{\eta\eta}$ gives $a_{\eta N}=(0.76 +i0.61)\,{\rm fm}$.

Therefore all the three versions of $t^{\eta\eta}$ have a pole
at $z=E_0-i\Gamma/2$, the first two of them reproduce the same
$a_{\eta N}$, the versions II and III are consistent with the
unitarity condition but give different $a_{\eta N}$.

In constructing the photoabsorption (production) $T$-matrix
$t^{\gamma\eta}$, we use the corresponding on-shell $T$-matrix
$t_{\rm on}^{\gamma \eta}(E)$ of Ref. \cite{Green99} and extend
it off the energy shell,
\begin{equation}
    t_{\rm off}^{\gamma \eta}(k',k;E) =
    \frac{\kappa^2 + E^2}{\kappa^2 + {k'}^2} \,
        t_{\rm on}^{\gamma \eta}(E) \,
    \frac{\alpha^2 + 2 \mu_{\eta N} E}{\alpha^2 + k^2}\ ,
\label{toff}
\end{equation}
using Yamaguchi form--factors which disappear (go to unity) on
the energy shell with $\kappa$ being a parameter.  Varying
$\kappa$ in our calculations, we found that the dependence of
the photoproduction cross-sections on the choice of this
parameter is rather weak, and we simply put $\kappa=\alpha$.  It
is known that $t^{\gamma \eta}$ is different for neutron and
proton.  In this work we assume that they have the same
functional form \eqref{toff} and differ by a constant factor,
$t_{\rm n}^{\gamma \eta} = A \, t_{\rm p}^{\gamma\eta}$.
Multipole analysis \cite{Muk} gives for this factor the
estimate: $ A= -0.84 \pm 0.15$\,.

To obtain the nuclear wave function $\psi_0$ (which is needed
for the expansion \eqref{expansion} of $H_A$), we solve the
few-body equations of the Integro-Differential Equation Approach
(IDEA)~\cite{idea} with the Malfliet--Tjon potential~\cite{mt}.

Figures 1 and 2 show the results of our calculations for the
total (integrated over the directions of the outgoing meson)
cross-section $\sigma$ of the coherent processes
\eqref{reactions}. The calculations were done for two nuclear
targets, namely, $^3$H and $^3$He. The curves corresponding to
the three versions of $t^{\eta\eta}$ are denoted respectively as
(I), (II), and (III).

\begin{figure}[h]
  \resizebox{16pc}{!}{\includegraphics{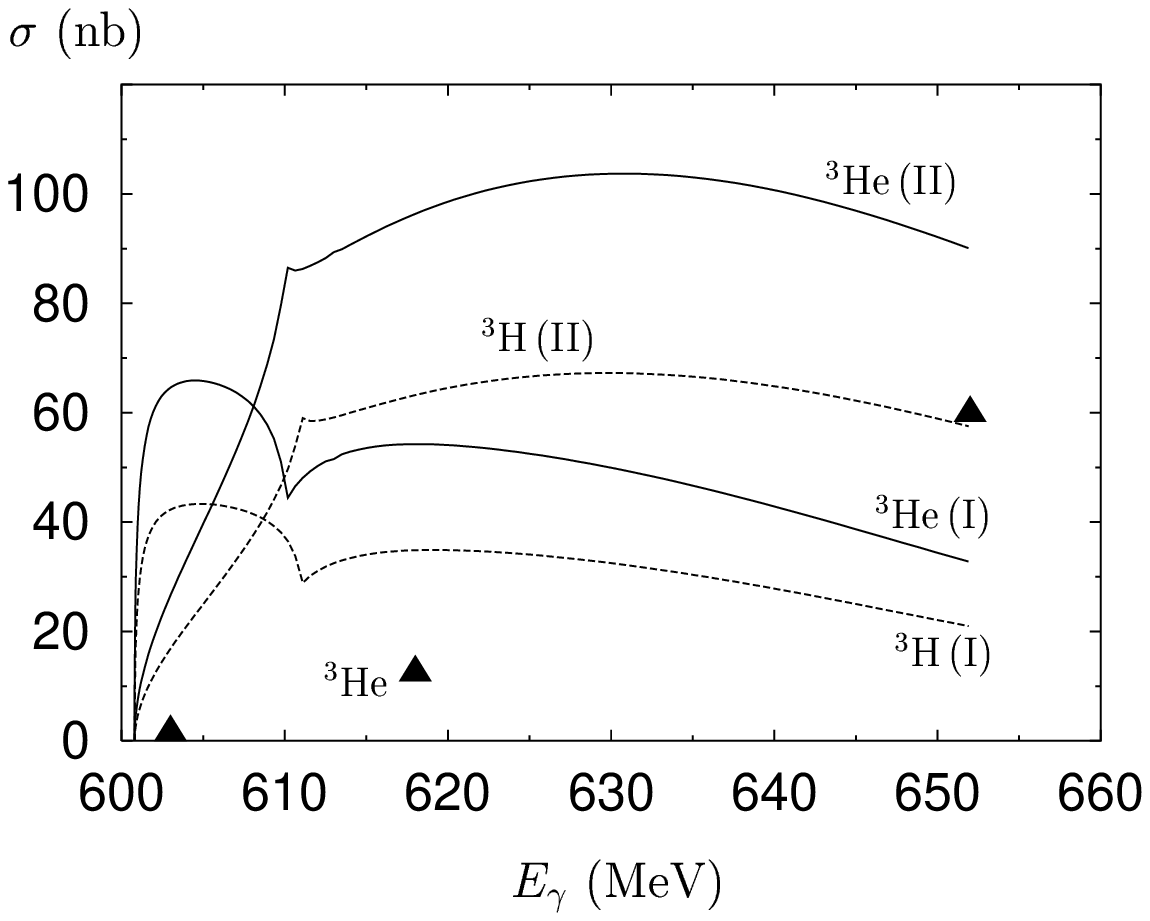}}
  \resizebox{16pc}{!}{\includegraphics{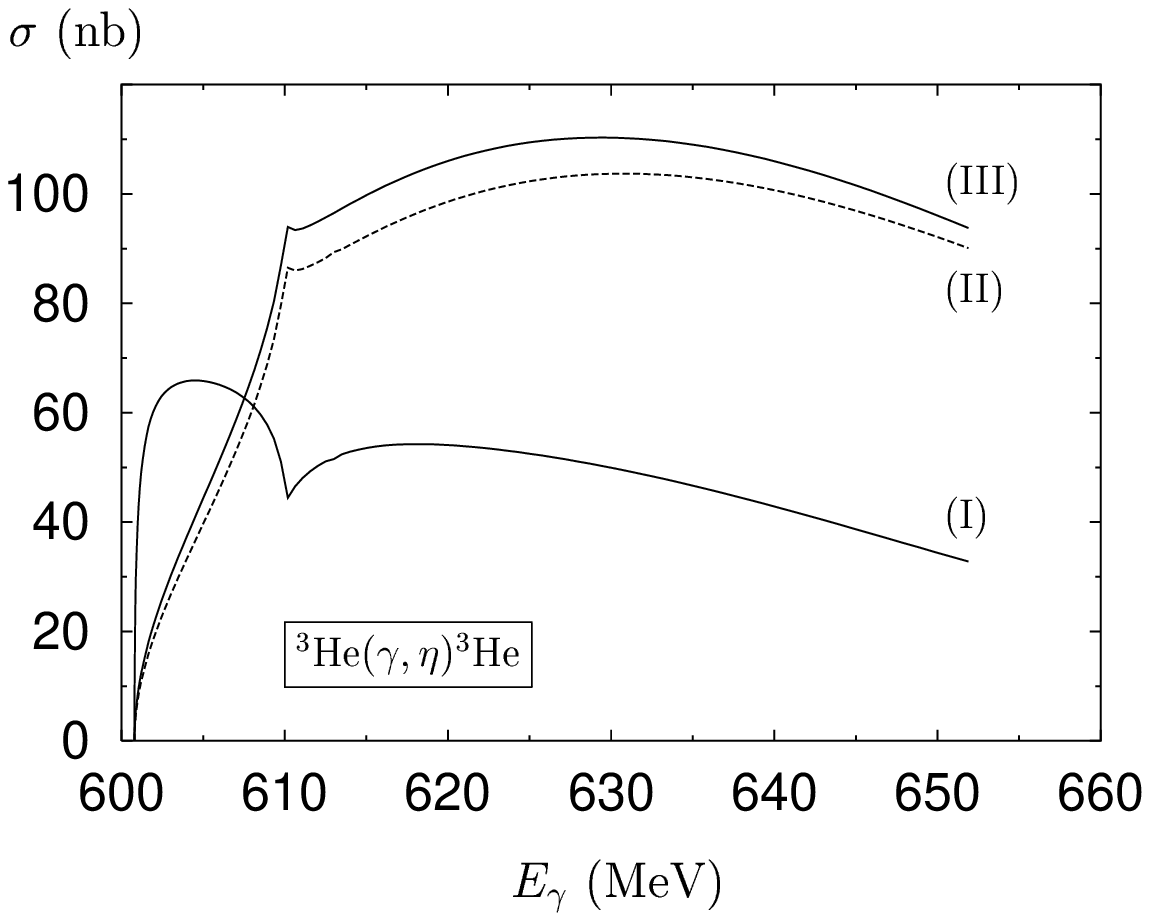}}
\caption{ Total photoproduction cross-section on ${}^3$He and
${}^3$H nuclei for different $t^{\eta \eta}$-matrices. Triangles
are the results for ${}^3$He taken from
Ref.\protect\cite{Tiator}. }
\label{fig1}
\end{figure}
As is seen in Fig.1 (left plot), the two versions of
$t^{\eta\eta}$, (I) and (II), give significantly different
results despite the fact that both of them reproduce the same
$a_{\eta N}$ and the $S_{11}$ resonance. This indicates that the
scattering of the $\eta$ meson on the nucleons (final state
interaction) is very important in the description of the
photoproduction process. This conclusion becomes even more
substantiated when our curves are compared to the corresponding
points (triangles) calculated for the $^3$He target in
Ref.~\cite{Tiator} where the final state interaction was treated
using an optical potential of the first order.  It is well-known
that the first-order optical theory is not adequate at the
energies near resonances. This is the reason why the
calculations of Ref.~\cite{Tiator} underestimate $\sigma$ near
the threshold.

\begin{figure}[ht]
  \resizebox{16pc}{!}{\includegraphics{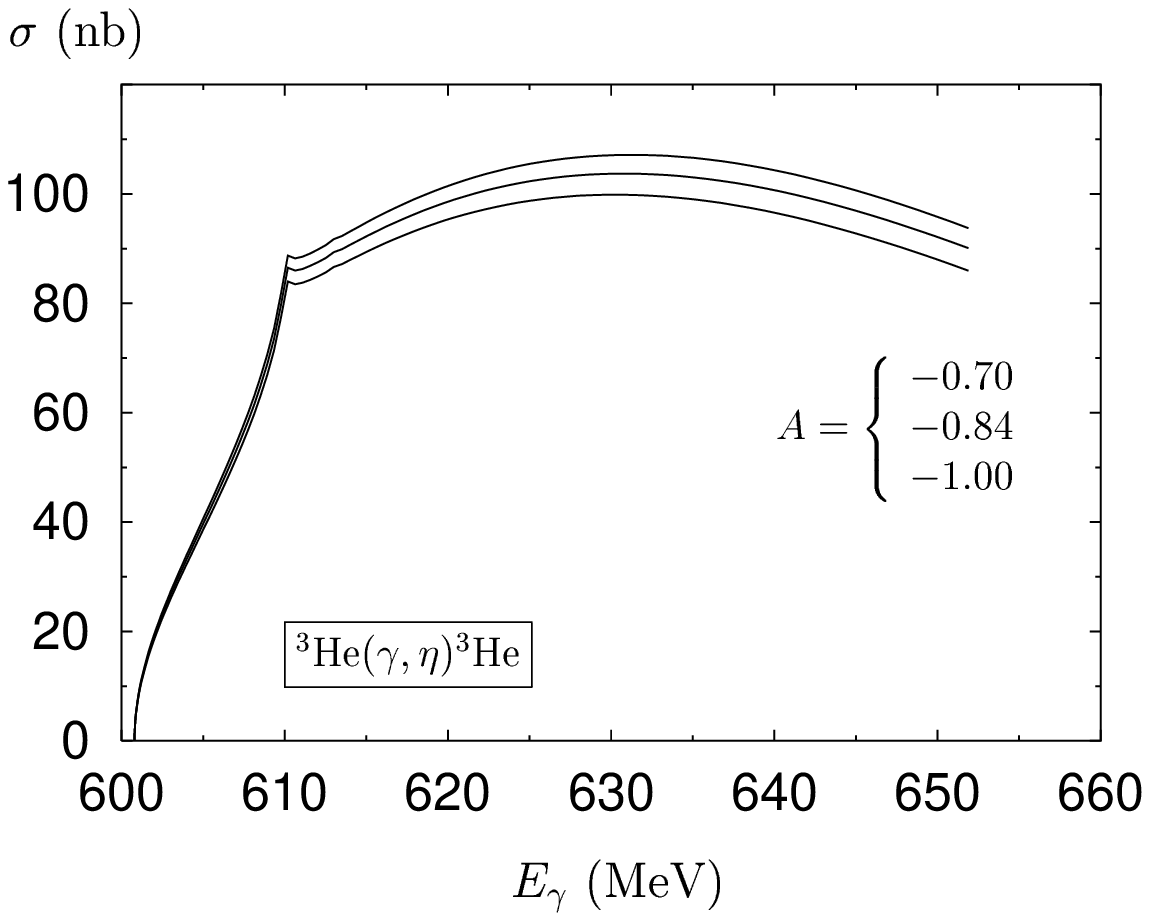}}
  \resizebox{16pc}{!}{\includegraphics{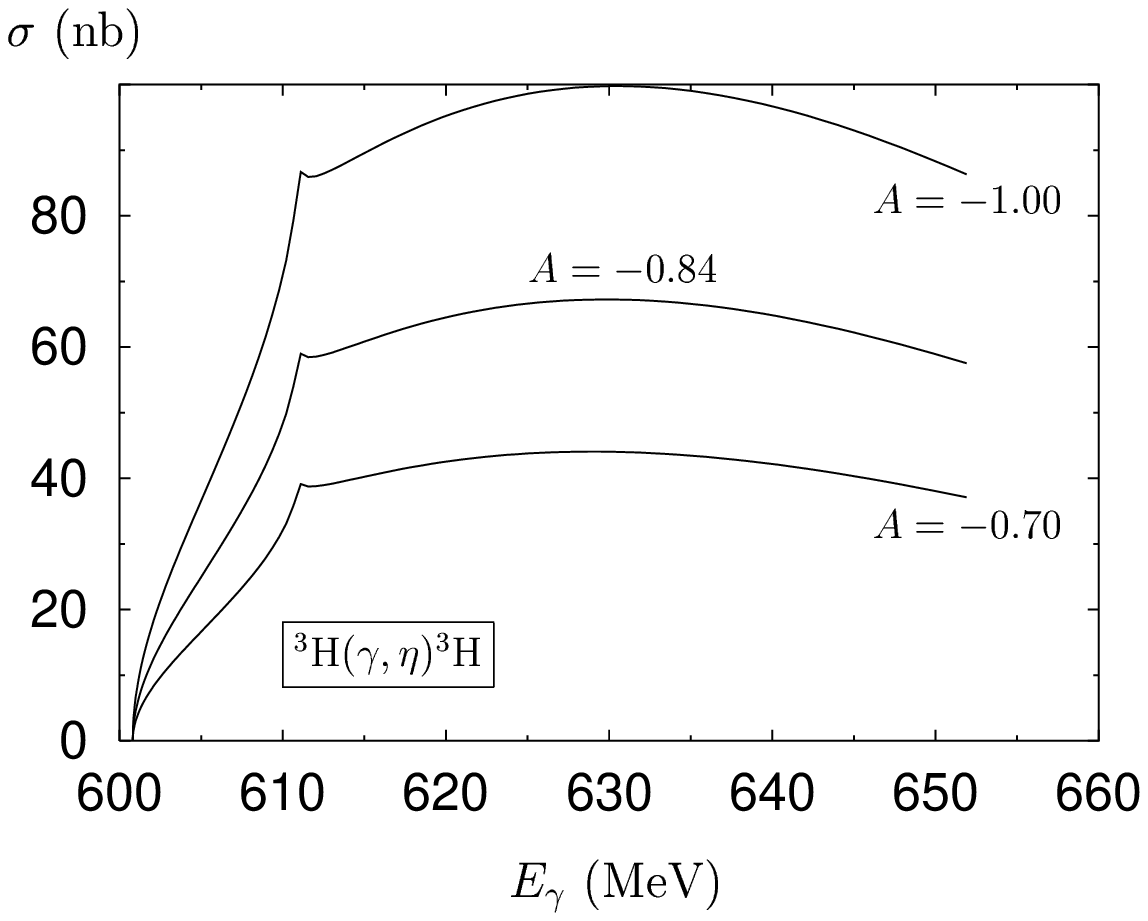}}
\caption{ Total photoproduction cross-section on ${}^3$He (left
plot) and ${}^3$H nuclei (right plot) for 3 different values of
parameter $A$. }
\label{fig2}
\end{figure}
Another conclusion, following from the fact that the curves (I)
in Fig.1 (left plot) are significantly different from the
corresponding curves (II), is that the two-body unitarity is
important as well. To clarify this statement, we compare (see
Fig.1, right plot) three curves corresponding to the three
choices of $\tau(z)$ in \eqref{tetan}. Surprisingly, the curves
(II) and (III) almost coincide despite the fact that they
correspond to different $a_{\eta N}$ while both obey the
two-body unitarity condition.

Fig.2 shows the dependence of $\sigma$ on the choice of the
parameter $A$ for the cases of $^3$He and $^3$H target (left and
right plots respectively). An interesting observation here is
that the cross-section for $\eta$ photoproduction is more
sensitive to this parameter when tritium rather than on $^3$He
target is used. This means that between these two nuclei, the
tritium is a preferable candidate for a possible experimental
determination of the ratio~$A$.

\vspace{5mm}
Authors would like to thank Division for Scientific
Affair of NATO for support (grant CRG LG 970110) and DFG-RFFR
for financial assistance (grant 436 RUS 113/425/1). One of
authors (V.B.B.) willing to thank Physikalishes Institut of Bonn
University for hospitality.


\end{document}